\documentclass[twoside,a4paper,11pt]{sea9}
\usepackage{graphicx}
\usepackage{hyperref}
\usepackage{movie15}
\usepackage[varg]{txfonts}

\begin{document}
\pagenumbering{arabic}
\pagestyle{myheadings}
\thispagestyle{empty}
{\flushleft\includegraphics[width=\textwidth,bb=58 650 590 680]{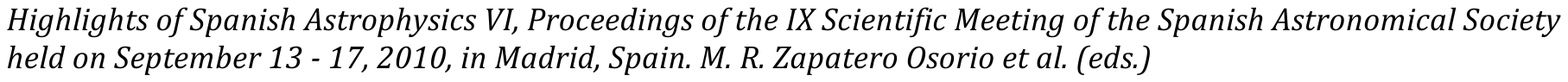}}
\vspace*{0.2cm}
\begin{flushleft}
{\bf {\LARGE
%
%%% TITLE of the paper. 
%%% TITLE of the paper. 
The coordinated key role of wet, mixed, and dry major mergers in the buildup of massive early-type galaxies at z$\lesssim$1
%
% Do not delete next few lines
}\\
\vspace*{1cm}
%
%%% Include here the LIST OF AUTHORS.
%%% Include here the LIST OF AUTHORS.
%%% Note that the last author has to be preceeded by an AND.
M.~Carmen Eliche-Moral$^{1}$, 
Mercedes Prieto$^{2,3}$, 
Jes\'us Gallego$^{1}$, 
and
Jaime Zamorano$^{1}$,
%
% Do not delete next few lines
}\\
\vspace*{0.5cm}
%
%%% AFFILIATIONS LIST.
%%% and the AFFILIATIONS LIST. Note that one affiliation per line.
%%% Add as many affiliations as necessary. 
$^{1}$
Departamento de Astrof\'{\i}sica, Universidad Complutense de Madrid, E-28040, Madrid, Spain\\
$^{2}$
Instituto de Astrof\'{\i}sica de Canarias, C/ V\'{\i}a L\'actea, E-38200, La Laguna, Tenerife, Spain\\
$^{3}$
Departamento de Astrof\'{\i}sica, Universidad de La Laguna, E-38200, Tenerife, Spain
%
% Do not delete next few lines
\end{flushleft}
%
% Headings
\markboth{
%%% Type the SHORT version of the paper title.
%%% Type the SHORT version of the paper title.
The role of wet, mixed, and dry major mergers in the buildup of mETGs since z$\sim$1 
}{ % Do not delete
%
%%%  First Author \& Second Author   OR   First-author et al.~
%%%  First Author \& Second Author   OR   First-author et al.~if the author list 
%%% contains three or more authors.
Eliche-Moral et al.% 
% Do not delete next few lines
}
\thispagestyle{empty}
\vspace*{0.4cm}
\begin{minipage}[l]{0.09\textwidth}
\ 
\end{minipage}
\begin{minipage}[r]{0.9\textwidth}
\vspace{1cm}
\section*{Abstract}{\small
%
% ABSTRACT ABSTRACT ABSTRACT
% ABSTRACT ABSTRACT ABSTRACT
%%% Type the ABSTRACT of your paper
Hierarchical models predict that massive early-type galaxies (mETGs) derive from the most massive and violent merging sequences occurred in the Universe. However, the role of wet, mixed, and dry major mergers in the assembly of mETGs is questioned by some recent observations. We have developed a semi-analytical model to test the feasibility of the major-merger origin hypothesis for mETGs, just accounting for the effects on galaxy evolution of the major mergers strictly reported by observations. The model proves that it is feasible to reproduce the observed number density evolution of mETGs since $z\sim 1$, just accounting for the coordinated effects of wet/mixed/dry major mergers. It can also reconcile the different assembly redshifts derived by hierarchical models and by mass downsizing data for mETGs, just considering that a mETG observed at a certain redshift is not necessarily in place since then. The model predicts that wet major mergers have controlled the mETGs buildup since z$\sim$1, although dry and mixed mergers have also played an essential role in it. The bulk of this assembly took place at 0.7$<z<$1, being nearly frozen at z$\lesssim$0.7 due to the negligible number of major mergers occurred per existing mETG since then. The model suggests that major mergers have been the main driver for the observational migration of mass from the massive end of the blue galaxy cloud to that of the red sequence in the last $\sim 8$\,Gyr.
%
% Do not delete next few lines
\normalsize}
\end{minipage}
%
%
%%% BODY of the paper
%%% BODY of the paper
%
\section{Introduction \label{sec:intro}}

Recent observations show that massive galaxies ($\log(\mathcal{M}_*/\mathcal{M}_\odot)\geq$11) have been in place since z$\sim$0.7 (\cite{2008ApJ...687...50P}). This epoch also coincides with the moment at which early-type galaxies (E-S0a's, ETGs) start to dominate the massive end of the galaxy luminosity function (LF), suggesting that there must exist "a dominant mechanism that links the shutdown of star formation and the acquisition of a spheroidal morphology in massive systems" (\cite{2010ApJ...709..644I}). According to current hierarchical models of galaxy formation, this mechanism has been major merging. The hierarchical scenario predicts that present-day mETGs have resulted from the most massive and violent merging sequences occurred in the Universe. Nevertheless, this also implies that they have been the latest systems to be in place into the cosmic scenario (at $z\sim 0.3$, see \cite{2006MNRAS.366..499D}), a prediction that conflicts directly with observations evidencing galaxy mass downsizing. Therefore, mETGs have turned into privileged laboratories to test hierarchical theories. However, studies on their physical properties have not provided conclusive results yet (see references in \cite{2010arXiv1003.0686E}), questioning or supporting the major-merger origin of mETGs (\cite{2006ApJ...644...54M,2010ApJ...718.1158L}). 

If mETGs derived mainly from major mergers, these galaxies should reside preferably in environments where mergers could have been relevant. Nevertheless, the evolutionary processes controlling the galaxy evolution in each environment are still poorly understood. In clusters, the S0 fraction is observed to increase with time at the expense of the spirals, but the fraction of ellipticals has remained constant since z$\sim$0.8 (\cite{2007ApJ...660.1151D,2009ApJ...697L.137P,2010arXiv1010.4442V}). This has been interpreted as a sign of the poor role played by major mergers in the evolution of cluster ellipticals, because it suggests the conversion of cluster spirals into S0's (but not into ellipticals) through mechanisms gentler than major mergers, contrary again to hierarchical expectations. However, the majority of present-day mETGs do not reside in clusters, but in groups (\cite{2006ApJS..167....1B}), an environment where mergers and tidal interactions determine galaxy evolution (\cite{2008ApJ...688L...5K}). In fact, the evolutive trend observed in groups is quite different to the one reported in clusters: old evolved groups evidence an evolution of spirals into S0's, and of these later into ellipticals with time  (\cite{2009AJ....138..295F,2010ApJ...713..637B}), suggesting a possible pre-processing of galaxies in groups before falling into clusters (\cite{2009AN....330.1059W}). So, the role played by the environment in galaxy evolution (and in particular, in the assembly of mETGs) is still unsettled. 

Mixed scenarios have been proposed for the formation of mETGs, in which blue galaxies quench their star formation through gas-rich mergers, migrate to the red sequence, and merge further through mixed and dry mergers to give place to the most massive ETGs (\cite{2007ApJ...665..265F}). Depending on the gas content of the encounter, the outcome of the major merger would be a S0 or an elliptical (\cite{2008ApJS..175..390H}). However, a direct verification of the feasibility of this scenario accounting strictly for the wet/mixed/dry major mergers reported by observations at each redshift has not been carried out yet. 

\section{The models \label{sec:models}}

We have approached this question directly through semi-analytical models, studying how the present-day mETGs would have evolved backwards-in-time assuming that they derive exclusively from the major mergers strictly reported by observations (\cite{2010A&A...519A..55E,2010arXiv1003.0686E}). We have accounted for the different phases, timescales, and progenitor galaxies of a major merger depending on its type (wet/mixed/dry), and for the number of major mergers of each type observed at each redshift. In particular, transitory phases in which gas-rich major mergers become dust-reddened star-forming galaxies (DSFs) have been considered. The model predictions apply to the global population of mETGs (averaged over all environments), as the local LFs and observational major merger fractions used in the model basically trace group and field environments. The effects of minor mergers have been neglected in the model (see comments concerning this topic in \cite{2010A&A...519A..55E}).

\begin{figure}[t!]
\center
\includegraphics[scale=.98]{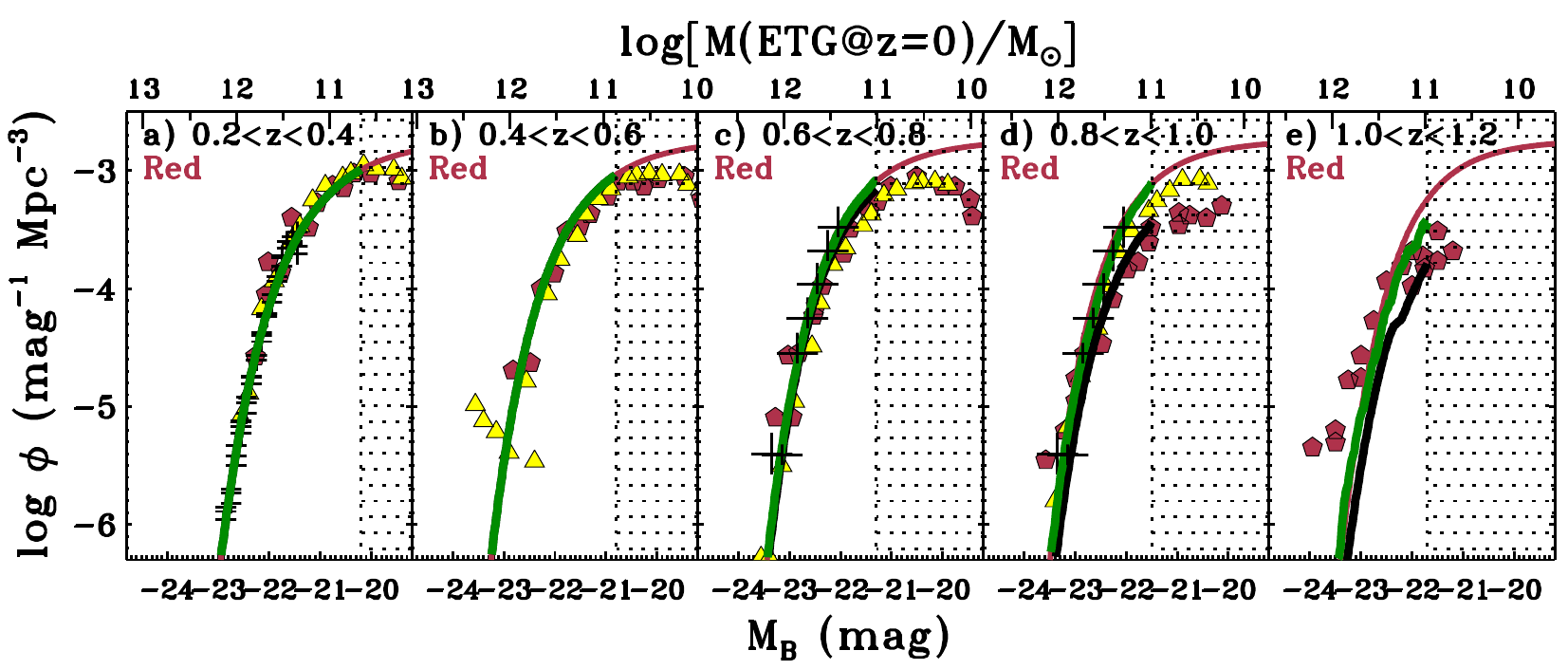} 
\includegraphics[scale=.98]{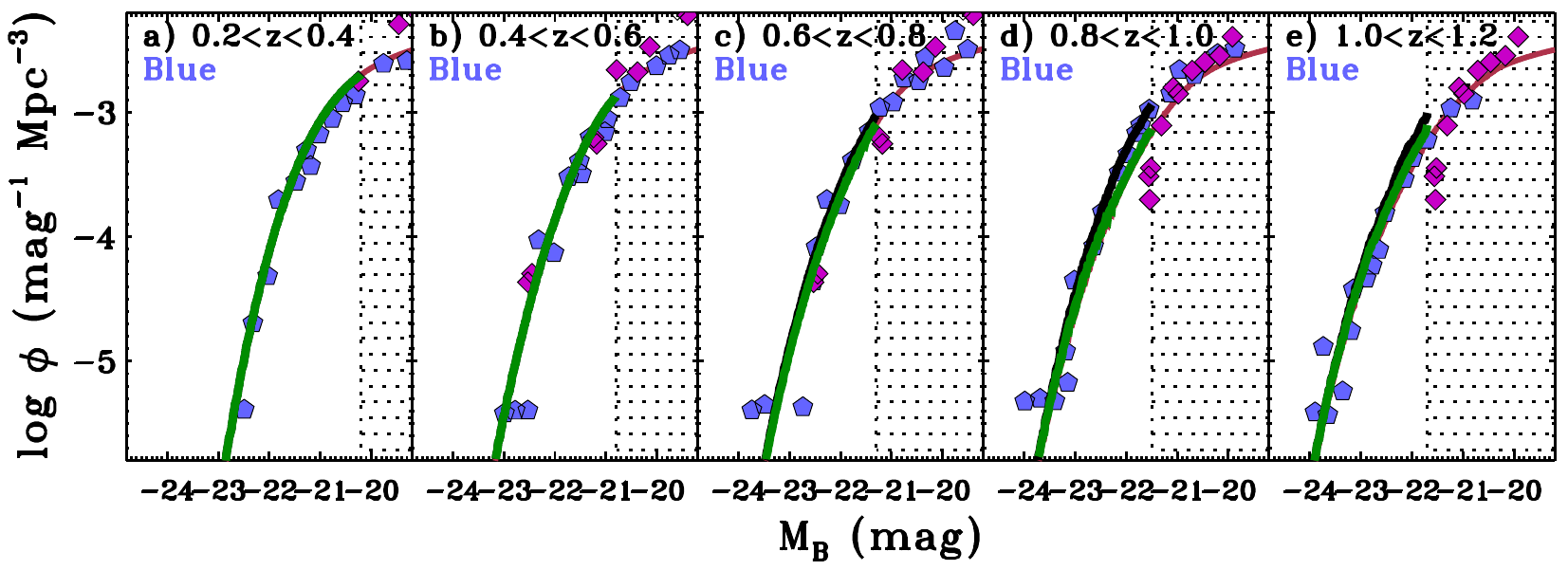} 
\caption{Evolution of the $B$-band LFs with redshift of red and blue galaxies up to $z\sim 1.2$. \emph{Symbols}: Observational data (see references in \cite{2010A&A...519A..55E}). \emph{Lines}: Model predictions. \emph{Black lines}: Model predictions considering that only ETGs are classified as red galaxies. \emph{Green lines}: Considering that also DSFs are identified as red galaxies (beside ETGs). \emph{Red thin lines}: Predictions of a pure luminosity evolution (PLE) model for all galaxies. \emph{Dotted shaded regions}: Magnitude range of model validity (more information in \cite{2010A&A...519A..55E}). Adapted from \cite{2010A&A...519A..55E}.
\label{fig:fig1}}
\end{figure}

\section{Reproducing the evolution of the galaxy LFs since z=1 \label{sec:observations}}

The model can reproduce the observed evolution of the galaxy LFs at z$\lesssim$1, simultaneously for different rest-frame bands ($B$, $I$, $K$) and selection criteria (based on color or morphology, see \cite{2010A&A...519A..55E}). Figure\,\ref{fig:fig1} shows the expected evolution of the LFs of red and blue galaxies in the $B$-band. The model predicts an increase of the number density of mETGs by a factor of $\sim$2-2.5 since z$\sim$1, in agreement with some studies (compare the black line and the pentagons in panel d of red galaxies). Moreover, if the DSFs predicted by the model are also considered as red galaxies, the model can also reproduce the apparent negligible number evolution of red galaxies reported by other authors (compare the PLE model, the model predictions including DSFs in the red sample, and the data -triangles and crosses- in the same panel). So, the model provides a framework in which conflicting observational results on the amount of number evolution experienced by massive red galaxies can be reconciled, just accounting for the typical contamination by DSFs that many red galaxy samples exhibit at $z>0.7$. The gas-rich progenitors of these "recently" assembled mETGs reproduce naturally the observational excess of $\sim$40\% of blue galaxies at 0.8$<z<$1, as compared to PLE models (see panel d of blue galaxies in the same figure). The model predicts a proportional rise in the number density of mETGs since z$\sim$1 similar for all mass bins (this also applies to mETGs with $\log(\mathcal{M}_*/\mathcal{M}_\odot)\gtrsim$11.7 at z$=$0). So, it is remarkable that, even predicting a noticeable buildup of the most massive ETGs (apparently in contradiction with mass downsizing), the LF of these systems predicted by the model at 0.8$<z<$1 lies on top of the PLE model (accordingly to observational data supporting mass downsizing). Consult \S\ref{sec:reconciling}. 

\begin{figure}[t!]
\center
\includegraphics[scale=1]{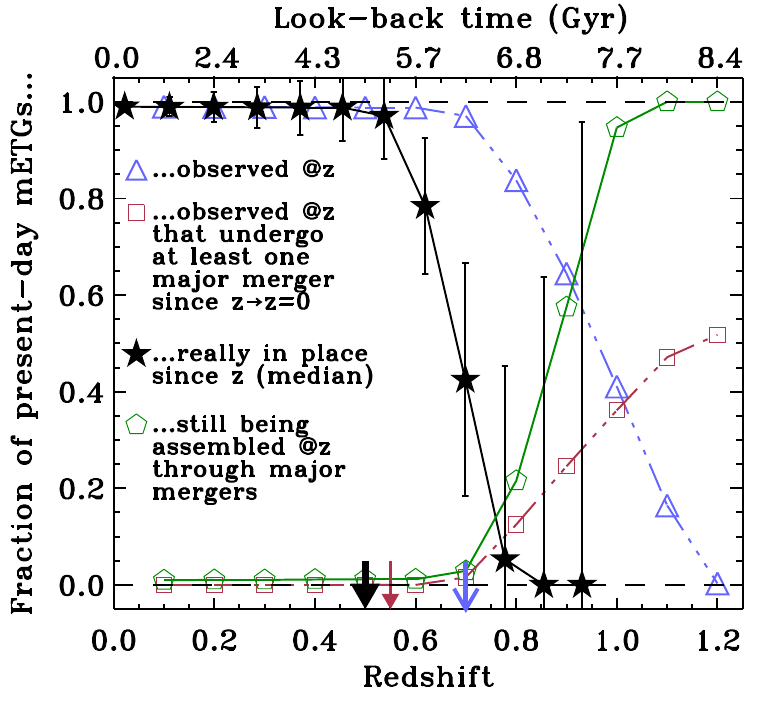} 
\caption{Model predictions on the fraction of present-day mETGs observed, really in place, and still being assembled at each redshift \emph{z}. \emph{Arrows}: Assembly redshift of mETGs according to observations (\emph{open blue}, \cite{2008ApJ...687...50P}), to the model (\emph{thick solid black}), and average $z$ between the assembly redshifts of 50\% and 80\% of their stellar content according to the models by \cite{2006MNRAS.366..499D} (\emph{red thin solid}). Our model reproduces the global observed buildup of mETGs since z$\sim$1 (triangles), predicting an assembly redshift for mETGs in good agreement with hierarchical models (\cite{2006MNRAS.366..499D}) at the same time. Adapted from \cite{2010arXiv1003.0686E}.
\label{fig:fig2}}
\end{figure}

\section{Reconciling the hierarchical scenario with mass downsizing\label{sec:reconciling}}

An ETG is in place since a given redshift $z$ basically if it does not undergo any major merger since then (not accounting for minor mergers, see \cite{2010arXiv1003.0686E}). This implies that a mETG observed at $z$ is not necessarily in place since then, as it depends on if it has undergone a major merger since then down to the present (contrary to what is usually assumed). Considering this, the model proves that mass downsizing and hierarchical scenarios can be reconciled easily. In fact, it predicts that more than $\sim$90\% of the mETGs existing at z$\sim$1 are going to be involved in a dry or mixed event since then (see Fig.\,\ref{fig:fig2}). This means that these mETGs are not the passively-evolved high-z counterparts of present-day ones (as usually interpreted in many studies), but their gas-poor progenitors instead. This implies that very few present-day mETGs have been really in place since z$\sim$1 ($\lesssim$5\%), contrary to the $\sim$40-50\% reported by traditional interpretations of observations. Accounting for this, the model is capable of deriving a lower assembly redshift for mETGs (z$\sim$0.5), in excellent agreement with hierarchical models, reproducing global mass downsizing trends at the same time (see Fig.\,\ref{fig:fig2}). 

\begin{figure}[t!]
\center
\includegraphics[scale=.95]{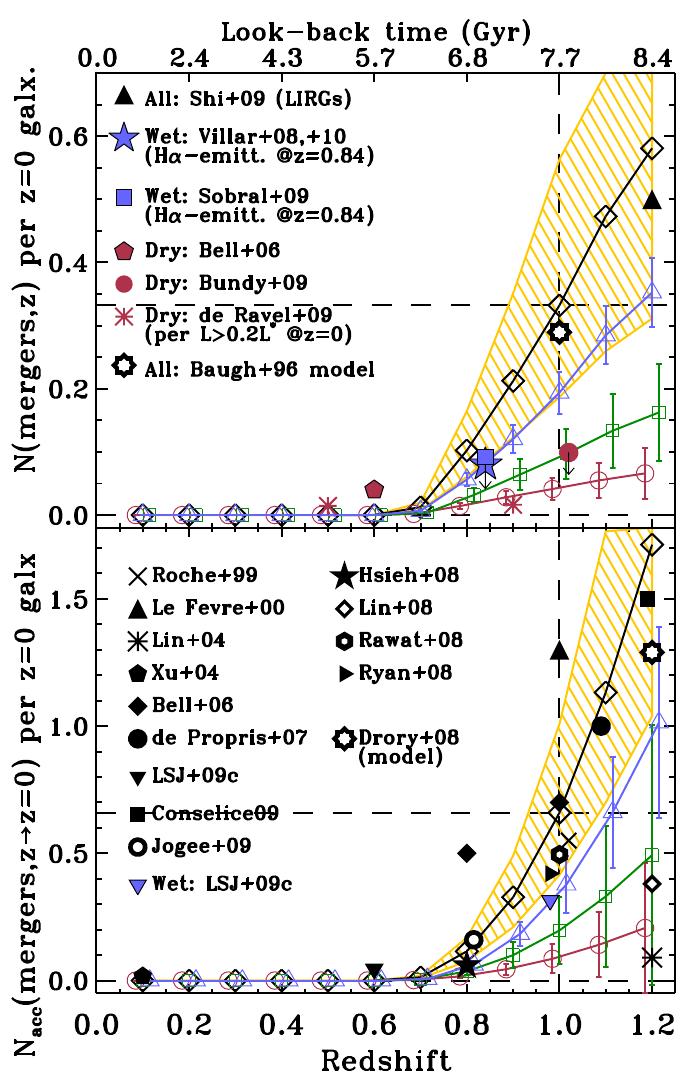} 
\includegraphics[scale=.95]{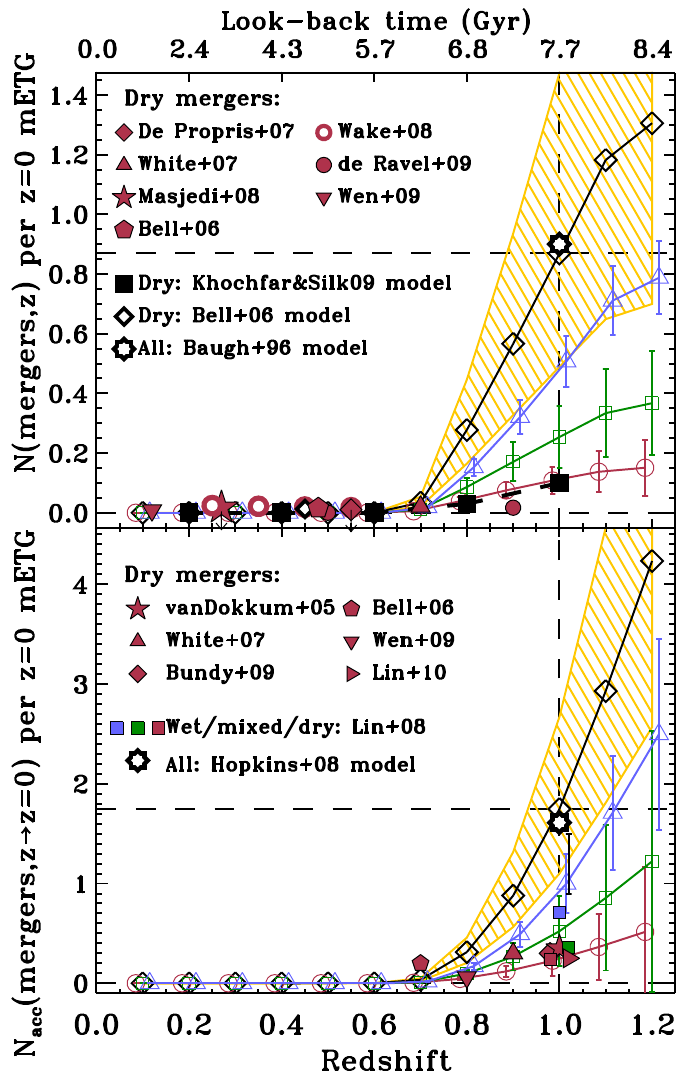} 
\caption{Predicted average number of wet, mixed, dry, and all major merger events occurred at each redshift bin $[$z-0.1,z$]$ and accumulated since each redshift $z$ down to the present, per present-day $L>L^*$ galaxy (\emph{top and bottom left panels, respectively}) and per local mETG (\emph{top and bottom right panels, respectively}). \emph{Open diamonds}: For all major mergers. \emph{Open triangles}: Wet mergers. \emph{Open squares}: Mixed mergers. \emph{Open circles}: Dry mergers. \emph{Remaining symbols}: Observational estimates derived by different studies (consult the legend in each panel). Adapted from \cite{2010arXiv1003.0686E}. 
\label{fig:fig3}}
\end{figure}

\section{The relative role of wet, mixed, and dry major mergers}

Accordingly to observations, the model predicts that wet major mergers have controlled the mETGs buildup since z$\sim$1, although dry and mixed mergers have also contributed significantly to it (see Fig.\,\ref{fig:fig3}). The bulk of this assembly took place during a $\sim$1.4\,Gyr time period elapsed at 0.7$<z<$1, being nearly frozen at z$\lesssim$0.7. This "frostbite" in the global buildup of mETGs is due to the negligible number of major mergers occurred since z$\sim$0.7, as compared to the mETGs population in place at those redshifts (see the right panels of Fig.\,\ref{fig:fig3}). Therefore, the contribution of major mergers (and, in particular, of dry events) to the global assembly of mETGs is predicted to have been irrelevant during the last $\sim$6.3\,Gyr. Notice that this is not necessarily extendable to clusters (as our models are biassed to field and group environments). The model predictions concerning to the numbers of wet/mixed/dry major mergers at each redshift are in excellent agreement with independent estimates (see Fig.\,\ref{fig:fig3}) and with observational studies also pointing to the fact that $z\sim 0.8$ is a transition epoch in the assembly of mETGs (\cite{2010ApJ...709..644I}). According to the model, major mergers are responsible for nearly doubling the stellar mass at the massive end of the red sequence since z$\sim$1, being dry events responsible of the accretion of $\sim$18\% of the stellar mass of present-day mETGs during the last $\sim$8\,Gyr. 
%
% Do not delete the next line
\small  % Do not delete
%
%%% Comment the following line if you do not have acknowledgments.
\section*{Acknowledgments}   % Do not delete if you declare acknowledgments
%
%%% ACKNOWLEDGMENTS
%%% ACKNOWLEDGMENTS
Supported by the Spanish Ministry of Science and Innovation (MICINN) under project AYA2009-10368 and by the Madrid Regional Government through the AstroMadrid Project (CAM S2009/ESP-1496). Funded by the Spanish MICINN under the Consolider-Ingenio 2010 Program grant CSD2006-00070: "First Science with the GTC".
%
% Do not delete the next few lines

%
\end{document}